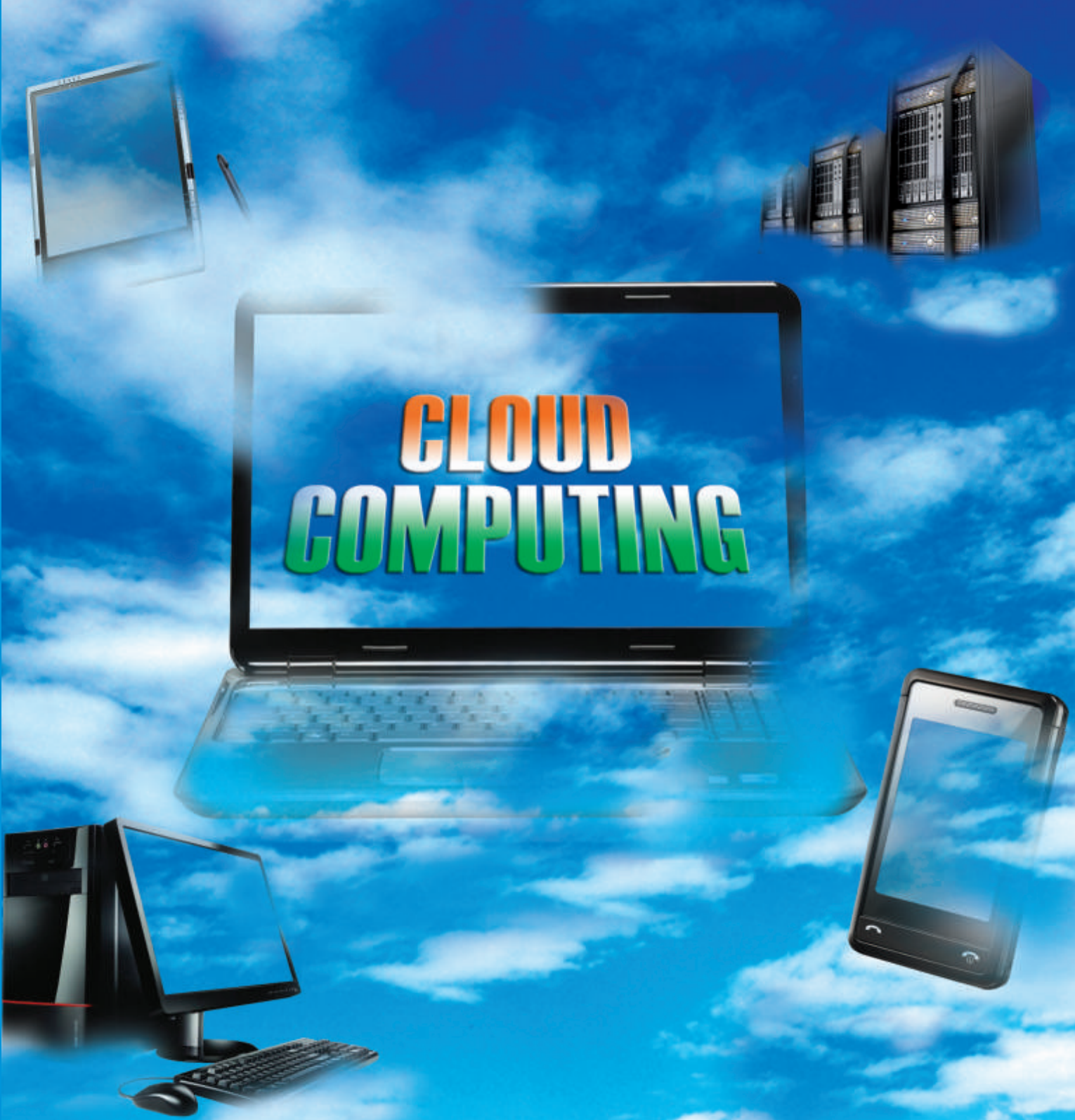

Cover Story

Rajkumar Buyya[1,2] and Karthik Sukumar[2]

[1] Cloud Computing and Distributed Systems (CLOUDS) Laboratory, Dept. of Computer Science and Software Engineering.
The University of Melbourne, Parkville, VIC 3010, Australia
[2] Manjrasoft Pvt. Ltd., ICT Building, 111, Barry Street, Carlton, Melbourne, VIC 3053, Australia. {karthik, raj}@manjrasoft.com

# Platforms for Building and Deploying Applications for Cloud Computing

Cloud computing is rapidly emerging as a new paradigm for delivering IT services as utlity-oriented services on subscription-basis. The rapid development of applications and their deployment in Cloud computing environments in efficient manner is a complex task. In this article, we give a brief introduction to Cloud computing technology and Platform as a Service, we examine the offerings in this category, and provide the basis for helping readers to understand basic application platform opportunities in Cloud by technology's such as Microsoft Azure, Sales Force, Google App, and Aneka for Cloud computing. We demonstrate that Manjrasoft Aneka is a Cloud Application Platform (CAP) leveraging these concepts and allowing an easy development of Cloud ready applications on a Private/Public/Hybrid Cloud. "Aneka CAP" offers facilities for quickly developing Cloud applications and a modular platform where additional services can be easily integrated to extend the system capabilities, thus being at pace with the rapidly evolution of Cloud computing.

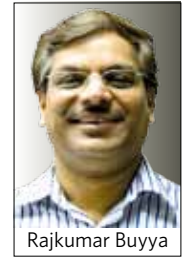
Rajkumar Buyya

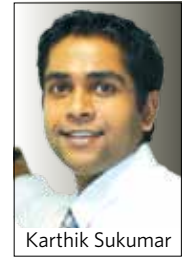
Karthik Sukumar

## 1. Introduction

Cloud computing is rapidly emerging as a new paradigm for delivering computing as a utility [1]. It allows leasing of IT capabilities whether they are infrastructure, platform, or software applications as services on subscription-oriented services in a pay-as-you-go model. Its foundation is based on various developments in IT during the last thirty to forty years. As fresh ideas and technology advancement have made it all the more striking and appealing during the Internet age, the way consumers consume and technology enablers deliver solutions has evolved. With a trend towards Cloud based model, the power is shifted to consumers. They have access to more compute power and to new applications, at an alluring price, as well as they enjoy the advantages of a self-service and self-managed environment.

Cloud computing fosters elasticity and seamless scalability of IT resources that are offered to end users as a service through Internet medium. Cloud computing can help enterprises improve the creation and delivery of IT solutions by providing them to access services in a most cost effective and flexible manner. A bird's eye view of Cloud computing is shown in Figure 1.

Although Cloud computing has emerged mainly from the appearance of public computing utilities [2], various deployment models, with variations in physical location and distribution, have been adopted. In this sense, regardless of its service class, Clouds can be classified as public, private, or hybrid depending on the model of deployment. A *public Cloud* is a Cloud made available in a pay-as-you-go manner to the general public. A *private Cloud* is a data center of an organization, not made available to the general public. A *hybrid Cloud* is a seamless use of public Cloud along with private Cloud when needed. In a typical public Cloud scenario, a third-party vendor delivers services such as computation, storage, networks, virtualization and applications to various customers. In a private Cloud environment, internal IT resources are used to serve their internal users and customers. Businesses are adopting public Cloud services to save capital expenditure and operational cost by leveraging Cloud's elastic scalability and market oriented costing features. Nevertheless, public Cloud computing also raises concerns about data security, management, data transfer, performance, and level of control.

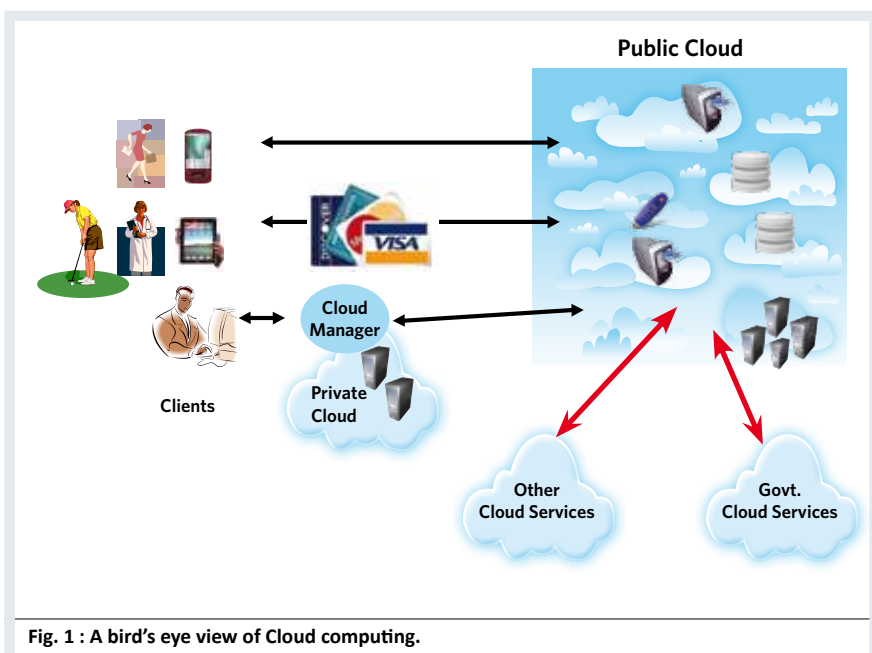

**Fig. 1 : A bird's eye view of Cloud computing.**

Cloud Computing started with a risk-free concept: let someone else take the ownership of setting up IT infrastructure and let end-users tap into it, paying only for what is been used. From this simple idea, a much more sophisticated, complex (and sometimes complicated) market started to grow. Today, businesses can buy computation resources, infrastructure plus platform or infrastructure plus applications. In the language of this market, the computation resources is frequently referred to as Infrastructure as a Service (IaaS), and the applications as Software as a Service (SaaS). In fact, use of the acronym appears ubiquitously from SaaS to PaaS (Platform as a Service) to XaaS (Anything as a Service). Key characteristics and vendors offering these Cloud services are highlighted in Fig.2.

What makes Cloud computing different from traditional IT approaches is the focus on service delivery and the consumer utilization model. In the background, service provider's uses particular technologies, system architecture, design and industry best practices to provide and support the delivery of service-oriented, elastically scalable environment serving multiple customers. This helps end users to have more agile and flexible service oriented architecture for their application and services. In a conventional IT scenario, most software companies have procured different components of their application middleware infrastructure layer from various vendors, and brought together these tools into a corporate environment using system integration services and tools. On the other hand, in a Cloud computing scenario, this practice is quite rare. Platform-as-a-Service solutions provide environment and applications development platforms for seamlessly integrating Cloud computing into existing application, services, and infrastructure with a market-oriented approach.

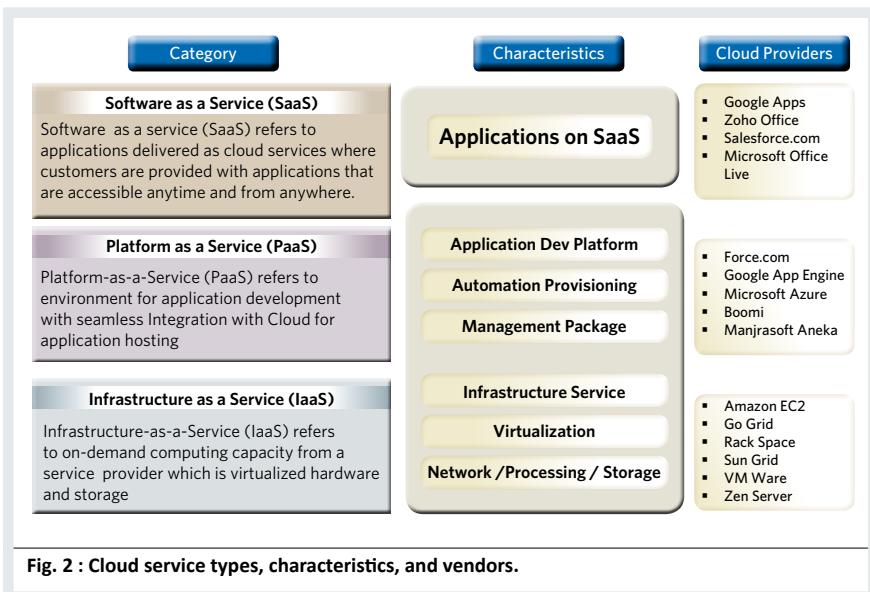

**Fig. 2 : Cloud service types, characteristics, and vendors.**

## 2. Cloud Application Development Platforms

Application development, deployment and runtime management have always been reliant on development platforms such as Microsoft's .NET, WebSphere, or JBoss, which have been deployed on-premise traditionally. In the Cloud-computing context, applications are generally deployed by Cloud providers to provide highly scalable and elastic services to as many end users as possible. The need for support as many users to access and utilize the same application services, with elastic resources allocation have led to enhancement in development platform technologies and architectures to handle performance, security, resource allocation, application monitoring, billing, and fault tolerance.

There are several solutions available in the PaaS market, to mention a few: Google App Engine, Microsoft Windows Azure, Force.Com, and Manjrasoft Aneka. Google App Engine provides an extensible runtime environment for web based applications developed with Java or Python, which leverage huge Google IT infrastructure. Windows Azure provides a wide array of Windows based services for developing and deploying windows based applications on the Cloud. It makes use of the infrastructure provided by Microsoft to host these services and scale them seamlessly. Aneka provides a more flexible model for developing distributed applications and provides integration with external Clouds such as Amazon EC2 and GoGrid. Aneka offers the possibility to select the most appropriate infrastructure deployment without being tied to any specific vendor–a virtual infrastructure, a private datacenter or a server –thus allowing enterprises to comfortably scale to the Cloud when needed.

### 2.1 Windows Azure

The Windows Azure Platform [3] consists of SQL Azure and the .NET services. The .NET services comprises of Access Control services and .NET service bus. Windows Azure is a platform with shared multitenant hardware provided by Microsoft. Windows Azure application development mandates the use of SQL Azure for RDBMS functionality, because that is the only coexisting DBMS functionality accessible in the same hardware context as the applications.

### 2.2 Google App Engine

Google App Engine is offered by Google Inc. Its key value is that developers can rapidly build small web based applications on their machine and deploy them on the Cloud. A notable thing is that Google App Engine provides developers with a simulated environment to build and test applications locally with any operating system or any system that runs a suitable version of Python and Java language environments. Google uses the Java Virtual Machine with Jetty Servlet engine and Java Data Objects.

### 2.3 Force.com

Force.com is a development and execution environment that is independent for Salesforce.com. Force.com is the best approach for Platform-as-a-Service (PaaS) for developing CRM based application and, with regards to the design of its platform and the runtime environment is based on the Java technology. The platform uses a proprietary programming language and environment called Apex code, which it has a reputation for simplicity in learning and rapid development and execution.

### 2.4 Manjrasoft Aneka

Aneka [4] is a distributed application platform for developing Cloud applications. Distributed means that Aneka can seam together any number of Windows based physical or virtual desktops or servers into a network of interconnected nodes that act as a single logical "application execution layer." The middleware is managed and monitored with advanced tools that allow monitoring applications' performance and the system status in order to meet the Service Level Agreements (SLAs) made with the users. Aneka-based Clouds can be deployed on a variety of hardware and operating systems including several flavors of the Windows and Linux operating system families. This flexibility allows Aneka to virtually harness almost all the different types of infrastructure and runtime environment to serve application execution on demand.

### 3. Aneka Cloud Application Platform

Aneka [4] is a platform for developing resource-intensive and elastic applications and their deployment on Clouds. It can harness a huge variety of physical and virtual resources, ranging from desktops, clusters, to virtual datacenters, to provide a single logical "application execution layer". The key components of the platform are depicted in Figure 3, which gives an overall view of Aneka from its foundations to the applications and the end user services. The platform is based on an extensible Service Oriented Architecture (SOA), which makes the integration of new components, incremental development of new features, and infrastructure deployment and configuration seamless tasks.

**Middleware.** The platform features a homogeneous distributed runtime environment for applications. Such environment is built by aggregating together physical and virtual nodes hosting the Aneka *container*. The container is lightweight layer that interfaces with the hosting environment and manages the services deployed on a node. Services constitute the core logic of Aneka Clouds and each container hosts three different classes of services:

- **Fabric Services**. Fabric services implement the fundamental operations of the infrastructure of the Cloud. These services include: high-availability and failover for improved reliability, node membership and directory, resource provisioning, performance monitoring and hardware profiling.
- **Foundation Services**. Foundation services constitute the core functionalities of the Aneka middleware. They provide a basic set of capabilities that enhance application execution in the Cloud. These services provide the infrastructure with added value and are both of use for system administrators and developers. Within this category we can list: storage management, resource reservation, reporting, accounting, billing, services monitoring, and licensing. Services in this level operate across all the range of supported application models.
- **Application Programming Services**. Application services deal directly with the execution of applications and are in charge of providing the appropriate runtime environment for each application model. At this level Aneka expresses its true potential in supporting different application models and distributed programming patterns. Aneka provides support for the most known application programming patterns such as distributed threads, bag of tasks, and MapReduce.

**Application Development and Management.** Aneka offers advanced features for developing and managing applications on the Cloud. The Software Development Kit (SDK) and the Management Kit are the two components exposing such capabilities. They provide means for interacting with the middleware and managing it with advanced user interfaces and bindings for applications. By using Aneka SDK, developers can quickly develop distributed applications, integrate the scaling capabilities of Aneka into existing applications, or implement new services to extend the potential of Aneka. The Management Kit allows deploying, managing, and tuning Aneka-based Clouds. By using a visual approach, it provides means to access and control every aspect of the middleware and also offers advanced features such as application reporting, accounting, billing, user management, and performance monitoring. The SDK and the Management Kit are the tools that enrich the user experience of developers and administrators respectively.

### 4. Cloud Application Programming Models and their Support in Aneka

Aneka Clouds aim to be a ubiquitous environment serving any type of computing need of distributed applications. Therefore, they are expected to be flexible enough to support several different models for developing applications: parameter sweep, concurrent, and data-intensive applications. In order to serve this purpose, Aneka provides engineers and developers with the concept of "Programming Model", which is collection of abstractions and runtime support for expressing and developing distributed applications. The platform currently supports three different programming models. They are: Task, Thread, and MapReduce [5]. Moreover, its extensible architecture offers the freedom to plug other models into the existing infrastructure. By taking this approach Aneka is able to provide support for all the following types of distributed applications:

**Parameter Sweep Applications: Task Programming.** Task Programming provides developers with the ability of expressing applications as a collection of independent tasks. Each task can perform a different operation, or the same operation on different data, and can be executed in any order by the runtime environment. Task programming allows the parallelization of legacy applications on the Cloud. Application from various domains including scientific computing, financial applications, media rendering and transcoding can be created using Task programming model. This model is the most popular in distributed computing and can be used as a starting point for implementing new models such as workflows with more complex requirements.

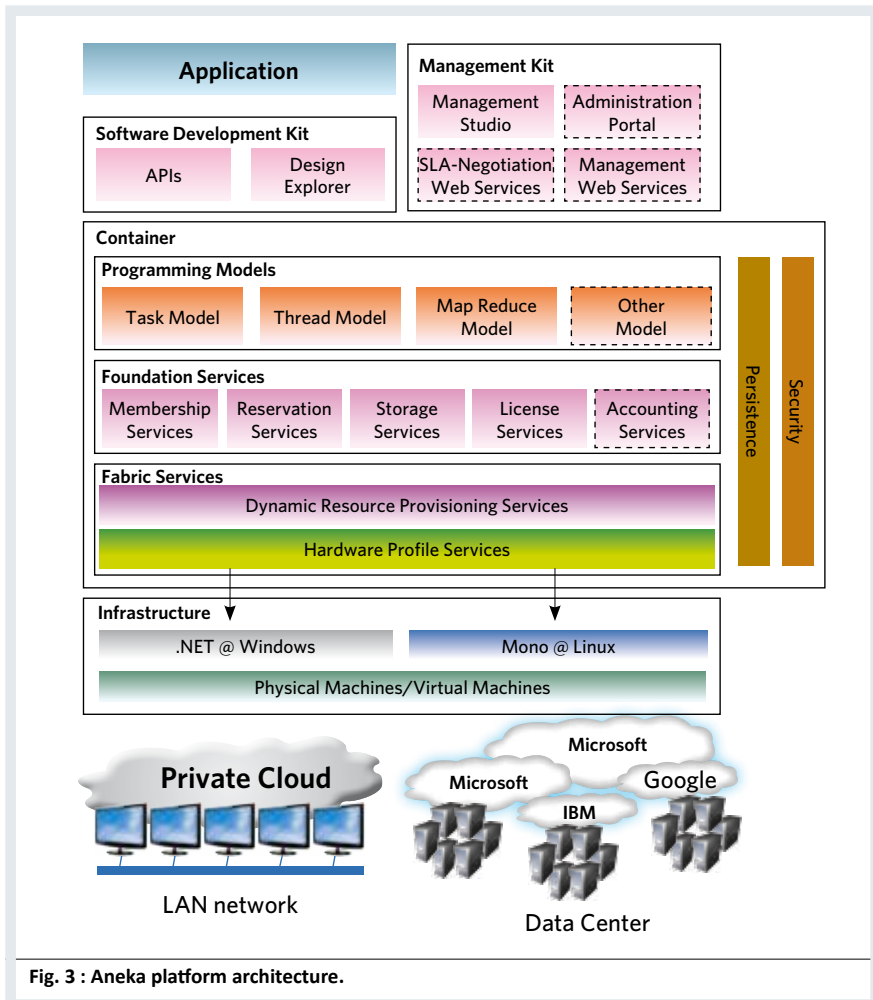

**Fig. 3 : Aneka platform architecture.**

**Concurrent Applications: Thread Programming.** The main abstraction of this model is the concept of *thread* which mimics the semantic of the common local thread but is executed remotely in a distributed environment. This model offers major control on the execution of the single components of an application but requires more management if compared with Task Programming, which is based on a "*submit and forget*" pattern. This model covers all the application scenarios of the Task programming and solves the additional challenges of providing a distributed runtime environment for local multi-threaded applications.

**Data Intensive Applications: MapReduce.** This model is an implementation of the MapReduce model, as proposed by Google, for .NET and integrated with Aneka. MapReduce has been designed to process a huge quantity of data by using simple operations that extracts useful information from a dataset (the map function) and aggregates this information together (the reduce function). MapReduce can be a winning solution for data mining and analytic applications, bulk media processing, and content indexing. Aneka provides a solid support for the model and integrates it with all the other foundation services such as accounting and reporting, thus making this solution a competitive alternative within the same market segment.

## 5. High-Performance Cloud Applications

Aneka has been used in creating several interesting applications in domains such as life sciences, engineering, and creative media. Applications created using Aneka are able to run on enterprise or public Clouds without any change. The three case studies on the use of Aneka for building applications in engineering, geospatial, and life science domains are discussed below.

### 5.1 Manufacturing and Engineering

The Manufacturing and Engineering sectors include a wide range of market segments, from aerospace to automotive. Manufacturing organizations face a number of computing challenges as they seek to optimize their IT environments, including high infrastructure costs and complexity to poor visibility into capacity and utilization. Today's design engineers need access to unrestrained, flexible computing capacity on demand, so that design cycles can be as fast, cheap, and productive.

The GoFront group, a division of China Southern Railway, is responsible for designing the high speed electric locomotive, metro car, urban transportation vehicle and the motor train. The raw design of the prototypes requires high quality 3D images using Autodesk's rendering software called Maya. By examining the 3D images, engineers identify problems in the original design and make the appropriate design improvements. However, such designs on a single four core served used to 3 days to render scenes with 2000 frames.

To reduce this time, GoFront has used Aneka and created an enterprise Cloud (see Figure 4) within their company by utilizing networked PCs. They used Aneka Design Explorer, a tool for rapid creation of parameter sweep applications, in which the same program is executed many times on different data items (in this case, executing the Maya software for rendering different images). A customized Design

Explorer (called Maya GUI) has been implemented for Maya rendering. Maya GUI managed parameters, generated Aneka tasks, monitored submitted Aneka tasks and collected completed rendered images. The design image used to take three days to render (2000+ frames, each frame with more than five different camera angles). Using only a 20 node Aneka Cloud, GoFront was able to reduce the rendering scenario from 3 days to 3 hours.

### 5.2 Geospatial Sciences and Technologies

Due to the continuous growth of GIS sciences and technologies, there have been even more geospatial and non-spatial data involved due to increase in number of data sources and advancement of data collection methodologies. Spatial analysis and Geo-computation are getting intricate and computationally demanding. The Department of Space, Government of India, adopted Aneka as the Cloud computing platform supporting the development of high performance GIS applications [6]. Aneka enables a new approach to complex analyses of massive data and computationally intensive environments, and gives the opportunity to satisfy all the requirements of a high-performance and distributed GIS environment over the public, private and hybrid Clouds.

### 5.3 Health and Life Science

With the high volume and density of data, along with the growing complexity of IT ecosystem and the pressures of competition and regulatory groups, life sciences organizations need IT infrastructure and management tools that can respond quickly to changing needs and, more importantly, enable rather than hamper the ability to innovate.

Aneka enables faster execution and massive data computation in life science R&D, clinical simulation, and business intelligence tools. It helps organizations to achieve greater levels of innovation in shorter timeframes while maximizing license utilization, increasing ROI, and realizing significant savings over Cloud based technology. For its application in real time scenario, Jeeva, an Enterprise Cloud enabled portal for protein secondary structure prediction, was developed based on Aneka. Research scientists use the portal to discover new prediction structures using parallel execution methods. The prediction took 20 minutes to complete when compared with the previous computational time of 8 hours. Also, Aneka has enabled implementation of personal health monitoring system aiding rehabilitation of stroke patients on public Cloud platform such as Amazon EC2.

### 5.4 IT Education and Research

As the IT field is rapidly moving towards Cloud Computing, software industry's focus is shifting from developing applications for PCs to Data Centers and Clouds that enable millions of users to make use of software simultaneously. This is creating a huge demand for manpower with skills in this area. Educational and research organizations require a platform that can support (1) multiple models of application programming, (2) multiple types of Cloud deployments (private, public, or hybrid), and (3) extensible framework enabling educators/researchers to develop their own programming models and application schedulers.

Since Aneka allows one to build a private Cloud with minimal investment by harnessing their existing IT resources (e.g., LAN-connected PCs), it has emerged as an excellent platform for teaching and research in Cloud computing area. A number of institutions in India such as MSRIT-Bangalore and C-DAC (Center for Development of Advanced Computing), Hyderabad, and other countries have used Aneka for setting up a Cloud Computing Lab and used it to offer practical exposure to their students (studying Cloud/Grid/High-Performance computing courses) in addition to building applications. Researchers in National Institute of Technology (NIT-Karnataka) have used Aneka for developing and evaluating new QoS-based resource provisioning and application scheduling algorithms. For case studies and to download Aneka platform, please visit http://www.manjrasoft.com/

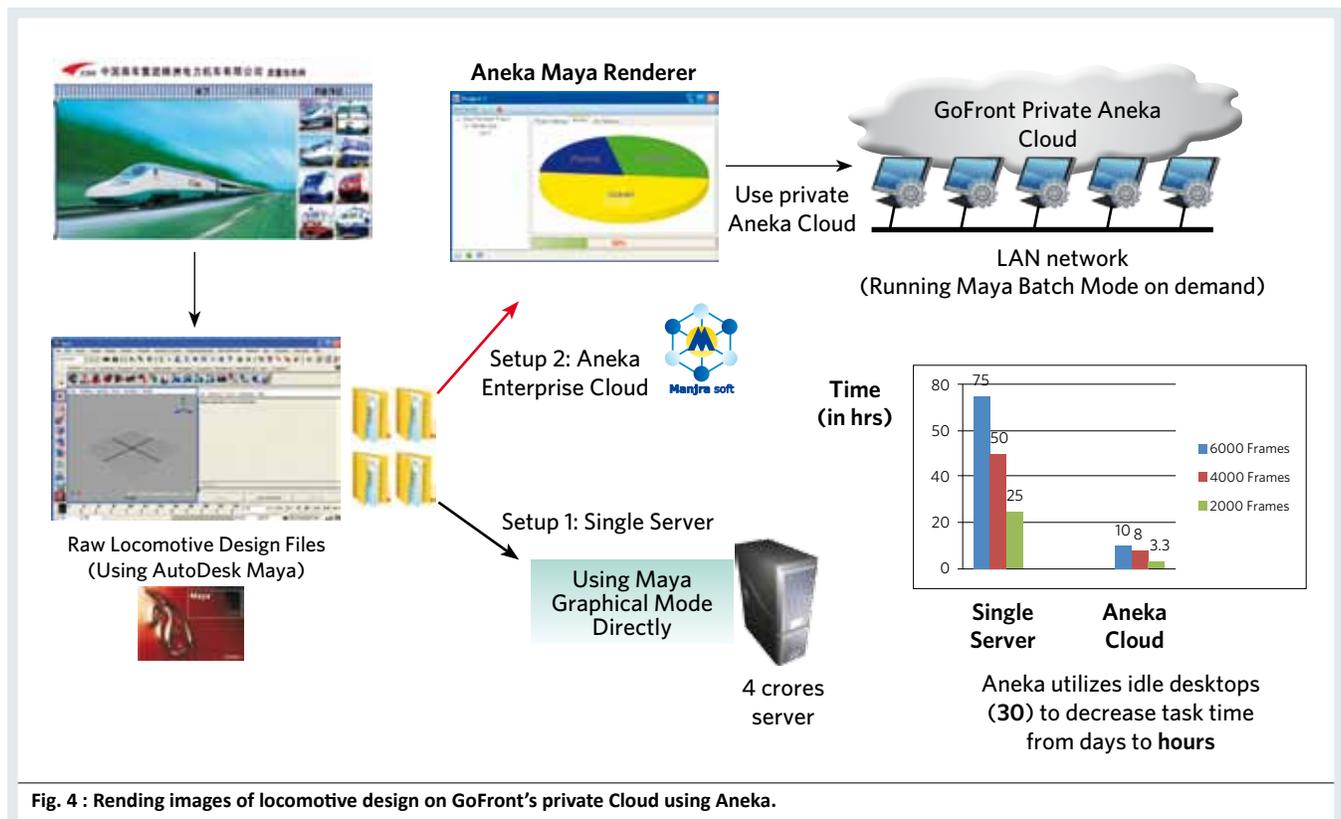

**Fig. 4 :** Rending images of locomotive design on GoFront's private Cloud using Aneka.

## 6. Conclusions and Future Directions

The growing interest in Cloud computing has led to new approaches for allocating financial resources and leveraging IT infrastructure and services. Cloud computing provides concrete opportunity for making a flexible use of IT by turning it into a utility. Cloud adoption is becoming a standard practice in many business sectors to scale IT infrastructure on demand. Despite this, the development of elastic and scalable applications is a complex task. Cloud application development platforms offer huge cost savings by reducing the cost of software engineering and enabling intelligent use of Cloud infrastructures. A wide range of applications scenarios from financial services, to entertainment and media, or manufacturing and engineering, demonstrates how Cloud technology can help increasing technology efficiency and adoption. PaaS technologies help organizations to harness their existing computing infrastructure and/or rent public Cloud infrastructure in a seamless manner. The true benefits of the Cloud application development will become apparent when developing and deploying application on solutions such as Aneka.

As the field of Cloud computing is rapidly progressing, there exist many opportunities for researchers and industrial developers to explore further. Key open issues that needs further investigation include: (1) Software Licensing, (2) Seamless integration of private and Cloud resources, (3) Security, Privacy and Trust, (4) Cloud "Lock-In" worries and Interoperability, (5) Application Scalability Across Multiple Clouds, (6) Clouds Federation and Cooperative Sharing, (7) Global Cloud Exchange and Market Maker, (8) Dynamic Pricing of Cloud Services, (9) Dynamic Negotiation and SLA Management, (10) Energy Efficient Resource Allocation and User QoS, (11), Power-Cost and $CO_2$ emission issues and seamless use of renewable and non-renewable electricity energy sources, and (12) Regulatory and Legal Issues.

## Acknowledgements

All members of CLOUDS Lab at the University of Melbourne and Manjrasoft have contributed towards various developments reported in this paper. In particular, we would like to thank Christian Vecchiola, Yi Wei, Xingchen Chu, Dileban Karunamoorthy, Suraj Pandey and Rodrigo Calheiros for their contributions towards the recent developments in Aneka and improving the paper. We thank Professor Geoffrey Fox whose seminar presentation in CLOUDS Lab has influenced on the content of this paper.All members of CLOUDS Lab at the University of Melbourne and Manjrasoft have contributed towards various developments reported in this paper. In particular, we would like to thank Christian Vecchiola, Yi Wei, Xingchen Chu, Dileban Karunamoorthy, Suraj Pandey and Rodrigo Calheiros for their contributions towards the recent developments in Aneka and improving the paper. We thank Professor Geoffrey Fox whose seminar presentation in CLOUDS Lab has influenced on the content of this paper.

## References

1. R. Buyya, C. Yeo, S. Venugopal, J. Broberg, and I. Brandic, *Cloud Computing and Emerging IT Platforms: Vision, Hype, and Reality for Delivering Computing as the 5th Utility*, Future Generation Computer Systems, 25(6):599-616, Elsevier, The Netherlands, June 2009.
2. R. Buyya, J. Broberg, and A. Goscinski (eds), *Cloud Computing: Principles and Paradigms*, Wiley Press, USA, Feb. 2011.
3. D. Chappell, *Introducing the Windows Azure Platform*, David Chappell & Associates, October 2010.
4. C. Vecchiola, X. Chu, and R. Buyya, *Aneka: A Software Platform for .NET-based Cloud Computing*, High Speed and Large Scale Scientific Computing, 267-295 pp., IOS Press, Amsterdam, Netherlands, 2009.
5. S Ghemawat and J Dean, *MapReduce: Simplified Data Processing on Large Clusters*, Proceedings of the 6th Symposium on Operating System Design and Implementation (OSDI'04), San Francisco, CA, USA, 2004.
6. K. Raghavendra, A. Akilan, N. Ravi, K. P. Kumar, and G. Varadan, *Satellite Data Product Generation Using Aneka Cloud*, Research Demo at the 10th IEEE International Symposium on Cluster, Cloud, and Grid Computing (CCGrid 2010), Melbourne, Australia, 2010.

### » About the Authors

**Dr. Rajkumar Buyya** is Professor of Computer Science and Software Engineering; and Director of the Cloud Computing and Distributed Systems (CLOUDS) Laboratory at the University of Melbourne, Australia. He is also serving as the founding CEO of Manjrasoft., a spin-off company of the University, commercializing its innovations in Cloud Computing. He has authored and published over 300 research papers and four text books. Software technologies for Grid and Cloud computing developed under Dr. Buyya's leadership have gained rapid acceptance and are in use at several academic institutions and commercial enterprises in 40 countries around the world. Dr. Buyya has led the establishment and development of key community activities, including serving as foundation Chair of the IEEE Technical Committee on Scalable Computing and five IEEE/ACM conferences. These contributions and international research leadership of Dr. Buyya are recognized through the award of "2009 IEEE Medal for Excellence in Scalable Computing" from the IEEE Computer Society, USA. Manjrasoft's Aneka Cloud technology developed under his leadership has received "2010 Asia Pacific Frost & Sullivan New Product Innovation Award".

**Karthik Sukumar** is serving as a Director of Manjrasoft, Australia in the area of product management and business analysis. He draws from a strong background in Software as a Service having served as a product manager at India's leading ISP, Sify Technology before joining Manjrasoft. He has performed and managed most of the roles of a modern software product organization and developed strategic partnership with leading organizations like IBM, Microsoft, Amazon, HP, VM Ware, and Citrix.